\documentclass[sigconf,natbib=false]{acmart}
\AtBeginDocument{%
  }

\setcopyright{none}
\acmYear{2026}
\acmConference[SAC'26]{The 41st ACM/SIGAPP Symposium on Applied Computing}{March 23--27, 2026}{Thessaloniki, Greece}
\thanks{\textcolor{red}{Preprint} accepted for publication in the Proceedings of the 41st ACM/SIGAPP Symposium on Applied Computing (SAC '26).}

\RequirePackage[
  datamodel=acmdatamodel,
  style=acmnumeric,
  ]{biblatex}

\usepackage{algorithmic}
\usepackage{algorithm}
\usepackage{array}
\usepackage[caption=false,font=normalsize,labelfont=sf,textfont=sf]{subfig}
\usepackage{textcomp}
\usepackage{stfloats}
\usepackage{url}
\usepackage{verbatim}
\usepackage{graphicx}

\usepackage{booktabs} 
\usepackage{makecell}
\usepackage{color}
\usepackage{hyperref}

\usepackage{listings}
\usepackage{xcolor} % For colors, optional but recommended

\lstdefinestyle{mystyle}{
    backgroundcolor=\color{black!5}, % light gray background
    basicstyle=\ttfamily\small,      % typewriter font, small size
    breakatwhitespace=true,          % break lines only at white space
    breaklines=true,                 % enable line breaking
    captionpos=b,                    % puts the caption at the bottom
    keepspaces=true,                 % keeps spaces in the text
    showspaces=false,                
    showstringspaces=false,
    showtabs=false,                  
    tabsize=2
}

\begin{document}

\title{LLM Chatbots in High School Programming: Exploring Behaviors
and Interventions}
% Please make sure that the short title does not exceed the width of one column
\renewcommand{\shorttitle}{LLMs in High School}

\author{Manuel Valle Torre}
\affiliation{%
  \institution{Delft University of Technology}
  \city{Delft}
  \country{The Netherlands}}
\email{m.valletorre@tudelft.nl}

\author{Marcus Specht}
\affiliation{%
  \institution{Delft University of Technology}
  \city{Delft}
  \country{The Netherlands}}
\email{m.m.specht@tudelft.nl}

\author{Catharine Oertel}
\affiliation{%
  \institution{Delft University of Technology}
  \city{Delft}
  \country{The Netherlands}}
\email{c.r.m.m.oertel@tudelft.nl}

% \author{Anonymous for Review}
% \affiliation{%
%   \institution{University}
%   \city{Ganymede}
%   \country{Jupiter}}
% \email{mail@mail.com}

% \author{Anonymous for Review}
% \affiliation{%
%   \institution{University}
%   \city{Ganymede}
%   \country{Jupiter}}
% \email{mail@mail.com}

% \author{Anonymous for Review}
% \affiliation{%
%   \institution{University}
%   \city{Ganymede}
%   \country{Jupiter}}
% \email{mail@mail.com}

%This command displays author info in page headers
% Please use the following convention:
% One author: J. Smith
% Two authors: J. Smith and I. Jones
% Three and more authors: J. Smith et al.
\renewcommand{\shortauthors}{Valle Torre et al.}
% \renewcommand{\shortauthors}{Anon et al.}

\begin{abstract}
This study uses a Design-Based Research (DBR) cycle to refine the integration of Large Language Models (LLMs) in high school programming education. 
The initial problem was identified in an Intervention Group where, in an unguided setting, a higher proportion of executive, solution-seeking queries correlated strongly and negatively with exam performance. 
A contemporaneous Comparison Group demonstrated that without guidance, these unproductive help-seeking patterns do not self-correct, with engagement fluctuating and eventually declining. 
This insight prompted a mid-course pedagogical intervention in the first group, designed to teach instrumental help-seeking. 
The subsequent evaluation confirmed the intervention's success, revealing a decrease in executive queries, as well as a shift toward more productive learning workflows. 
However, this behavioral change did not translate into a statistically significant improvement in exam grades, suggesting that altering tool-use strategies alone may be insufficient to overcome foundational knowledge gaps. 
The DBR process thus yields a more nuanced principle: the educational value of an LLM depends on a pedagogy that scaffolds help-seeking, but this is only one part of the complex process of learning.
\end{abstract}




\maketitle

\section{Introduction}
% \textcolor{red}{Symposium On Applied Computing (ACM SAC 2026): Artificial Intelligence for Education, 8 pages, https://easychair.org/cfp/SAC2026}

As technology becomes increasingly integral to education and society, the need for widespread computational skills and programming grows \cite{becker_programming_2023}. 
Learning to program, however, presents unique challenges: 
Students frequently struggle with abstract concepts such as loops and conditionals, find error messages cryptic, and may default to memorizing syntax rather than understanding programming logic \cite{cheah_factors_2020}.
These difficulties often lead to frustration, reduced motivation, and high dropout rates, highlighting a critical need for timely, individual support \cite{xie_developing_2023,price_factors_2017}.

Recent advancements in Large Language Models (LLMs) present promising solutions for overcoming these educational barriers \cite{giannakos_promise_2024}. 
By providing personalized, real-time assistance, LLMs have the potential to help students better manage cognitive load, clarify confusing syntax, and focus more deeply on programming concepts \cite{park_empowering_2024}. 
At the same time, the integration of LLMs in educational settings carries risks.
Over-reliance on AI assistance may weaken students' independent problem-solving skills, encourage superficial learning strategies, and even hinder the development of key cognitive skills like critical thinking when compared to non-AI approaches \cite{zhai_effects_2024,fan_beware_2024,ji_humanmachine_2025}. 
Furthermore, potential inaccuracies or misleading responses from LLMs raise concerns about their reliability as educational assistants \cite{giannakos_promise_2024}.

While the importance of guiding student use of GenAI is recognized and there are opportunities to foster metacognitive processes \cite{zhang_systematic_2024}, there is a lack of research on structured interventions designed to promote effective use of LLMs. 
Some prior work has explored data-driven guidance systems for help features in tutoring systems \cite{marwan_unproductive_2020}, but applying these ideas to the context of GenAI requires further investigation. 
Moreover, most studies focus on university-level learners, leaving high school students less explored \cite{ng_empowering_2024}. 
In short, more empirical research is needed to understand the effects and underlying mechanisms of LLM use in education \cite{wu_ai_2024}.

To address these gaps, this study employs a Design-Based Research (DBR) cycle to investigate and refine the integration of LLMs in a high school programming course. 
The study is centered on an Intervention Group (IG), where we first identified a core problem: in an unguided setting, a higher frequency of LLM use—particularly for executive, solution-seeking queries—negatively correlated with midterm exam performance \cite{aleven_help_2016}. 
This critical insight informed the design of a targeted pedagogical intervention, which was then implemented and evaluated in the second phase of the course. 
Data from a contemporaneous Comparison Group (CG) provides additional context for our findings.

The following research questions guide this study:
\begin{itemize}
    \item \textbf{RQ1 - Baseline}: How does the frequency of student interactions with an LLM correlate with their programming performance?
    \item \textbf{RQ2 - Baseline}: How do instrumental vs. executive query types impact learning outcomes in an unguided setting?
    \item \textbf{RQ3 - Intervention}: How does a targeted intervention influence students' help-seeking behaviors and learning outcomes?
\end{itemize}
The following sections build the case for our questions by reviewing prior work, detailing our research design, presenting the results of our analysis, and discussing the implications for supporting high school programming learners.

\begin{figure*}[b]
    \centering
    \includegraphics[width=0.8\textwidth]{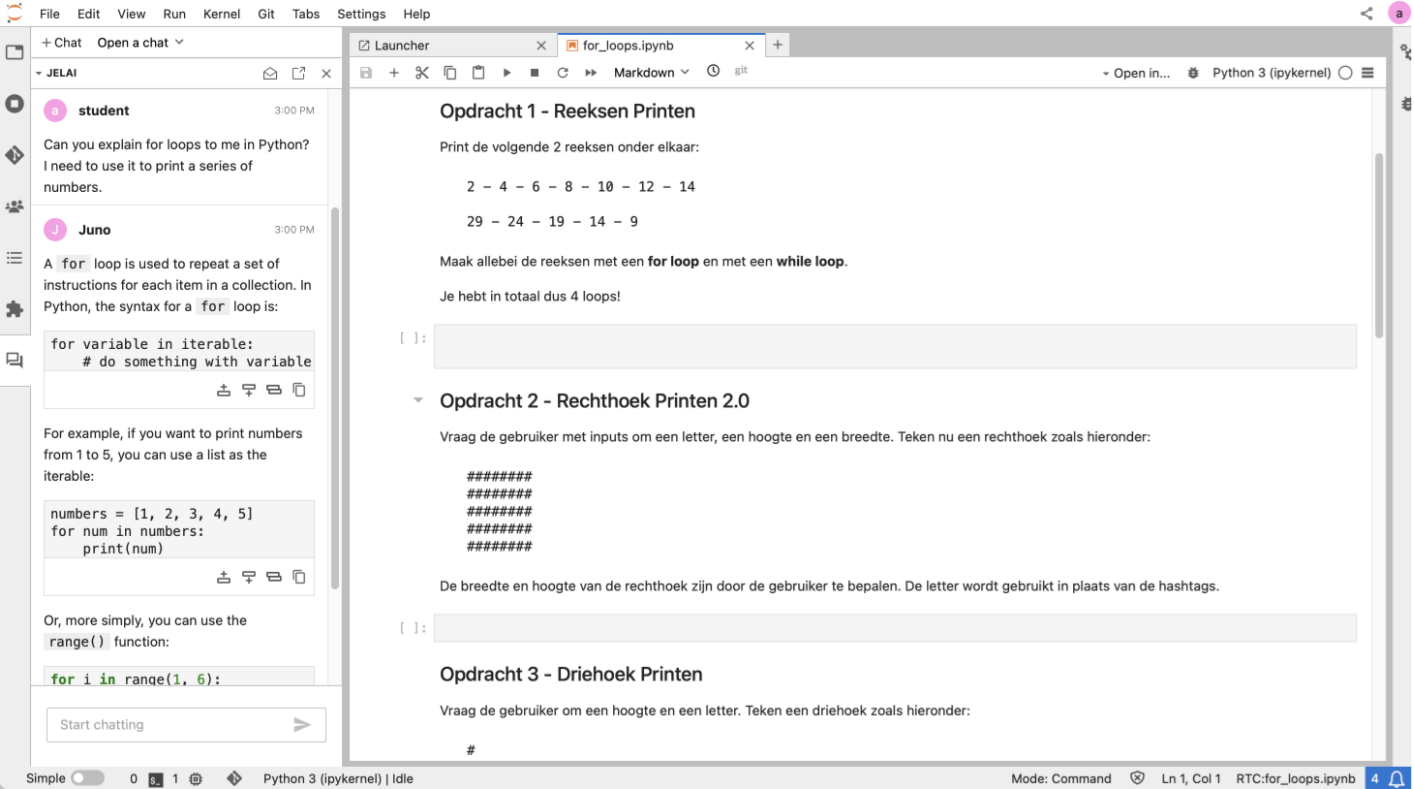}
    \caption{JELAI interface with a sample task in the Jupyter Notebook on the right and a task-specific LLM-based tutor on the left}
    \label{fig:JELAI}
\end{figure*}

\section{Related Work}
Effective pedagogical support extends beyond providing direct solutions; it also involves guiding students toward appropriate help-seeking strategies.
Help-seeking theory offers a valuable framework for understanding how students request and utilize assistance, which is highly relevant when introducing AI helpers \cite{karumbaiah_context_2022}. 
Help-seeking has been framed as a positive, strategic part of learning, rather than a sign of failure, and distinguished between executive and instrumental help-seeking \cite{nelson-le_gall_help-seeking_1981}.

Executive help-seeking is considered non-adaptive; the student aims to minimize effort by having someone else (or an AI) complete the task, for example, by asking, “Can you just give me the answer to this problem?”
In contrast, instrumental help-seeking is an adaptive, learning-focused strategy where the student requests just enough help (e.g., a hint or explanation) to overcome an impasse and move forward independently \cite{ko_trees_2024}. 
Instrumental help-seeking is strongly associated with better retention and skill transfer \cite{roll_benefits_2014}. 
Ultimately, help-seeking behavior is often considered a key aspect of the broader construct of self-regulated learning (SRL) \cite{karumbaiah_context_2022}. 

This active role is vital for learning, as even well-crafted instructional explanations can be ineffective if the learner does not engage with them to develop their own understanding \cite{wittwer_why_2008}. 
By asking instrumental questions, students are prepared to engage in cognitive activities, such as generating inferences or revising their mental models. 
This process helps them integrate new information with prior knowledge, thereby constructing a more robust understanding of the material \cite{kozanitis_perception_2007}.

LLM-based tutors make this process even more relevant. 
Whereas prior research on ITS often studied students requesting incremental hints \cite{aleven_example-tracing_2016}, LLMs typically provide complete explanations, akin to bottom-out hints \cite{pal_chowdhury_autotutor_2024}. 
This fundamental shift from minimal guidance to comprehensive answers makes it essential to reevaluate whether established help-seeking principles remain, and if students can be guided towards more productive learning behaviors \cite{ng_empowering_2024}.

% \section{Related Work}
\subsection{Programming Education and Large Language Models}
Novice programmers must simultaneously learn new syntax and logic while also interpreting error messages \cite{yang_systematic_2023,zambach_ai-enhanced_2025}. 
Studies comparing novice and expert coders highlight these differences, showing novices make more errors and struggle to achieve program correctness \cite{chuang_analyzing_2024}. 
Effective help-seeking behaviors are crucial for overcoming these obstacles, yet novices often lack the metacognitive skills or confidence to ask for appropriate assistance \cite{roll_benefits_2014,karabenick_understanding_2011,marwan_unproductive_2020}. 
These challenges underscore the need for timely, individualized, and scalable educational support.

LLMs offer scalable, personalized support in programming education, potentially replicating the benefits of human tutors by explaining code and reducing stress.
Recent advancements of LLMs offer potential solutions for providing this in programming education \cite{giannakos_promise_2024,park_empowering_2024}. 
LLMs can generate and explain code, answer questions immediately, potentially reduce stress, and improve task efficiency \cite{sheese_patterns_2024}. 
Furthermore, some students may feel more comfortable asking basic or repetitive questions to an LLM compared to a human instructor \cite{park_empowering_2024,hou_effects_2024}.

However, the integration of LLMs like ChatGPT also presents significant challenges and risks \cite{giannakos_promise_2024}.
Firstly, these tools can complicate the programming learning experience, giving confusing or advanced information \cite{tossell_student_2024}.
If students fail to see the usefulness of these systems, they may stop using them before getting any benefits from them \cite{setala_use_2025}.
Standard LLMs are not designed as pedagogical tools; they typically prioritize task completion over fostering deeper conceptual understanding \cite{zhai_effects_2024,fan_beware_2024}. 
This can lead to superficial engagement and potentially hinder the development of independent problem-solving and debugging skills \cite{sheese_patterns_2024}. 

Recognizing these complexities, researchers have developed specialized LLM-based systems for programming education, aiming to provide help that supports learning.
Some prominent examples are:
\textbf{CodeTutor}: A chatbot using system prompts for educational responses; its interface is similar to that of ChatGPT and allows students to rate feedback at both the message and conversation levels \cite{lyu_evaluating_2024}.
\textbf{CodeAid}: An AI tutor providing targeted support, guiding the student via specific feature selection while avoiding direct solutions \cite{kazemitabaar_codeaid_2024}.
And \textbf{CodeHelp}: A web interface using multiple prompts to ensure an appropriate response from the LLM, which also avoids giving away solutions \cite{sheese_patterns_2024}.

While these systems represent important advances, the studies evaluating them often rely on post-task surveys or analysis of conversations in a separate chat window, rather than logging interactions that occur seamlessly within the coding environment itself. 
The system used in this study, JELAI, is designed to address this by integrating the LLM chatbot directly into the notebook environment, similar to modern development interfaces (like GitHub Copilot or Cursor), potentially making interaction more intuitive.

\section{Methodology}

\subsection{Research Design and Procedure}
Integrating LLMs into authentic classroom settings presents a complex, ill-defined problem for which Design-Based Research (DBR) is an ideal methodological framework. 
DBR enables us to move beyond efficacy questions ("does it work?") to develop a deeper, theoretically grounded understanding of how, why, and under what conditions an intervention functions in a real classroom. 
Our study follows a classic DBR cycle: we first used help-seeking theory to analyze an existing implementation and identify a core pedagogical problem, then designed and implemented a targeted intervention to address it, with the ultimate goal of generating shareable design principles.
The aim of this DBR cycle, therefore, is not statistical generalization to a wider population, but rather the development of a local theory and shareable design principles grounded in the complex, authentic context of this specific classroom.

The design centers on a pre-post analysis of an Intervention Group (IG, N = 18) and a non-equivalent Comparison Group (CG, N = 19) to contextualize the impact of the intervention. 
We maintained authentic class sizes to preserve ecological validity and avoid confounding variables (e.g., teacher effects) associated with merging disparate cohorts.

Participants in both groups were recruited from elective Informatica classes at two different Dutch high schools, where they were introduced to foundational programming concepts (e.g., variables, conditionals, loops). 
The non-equivalent Comparison Group was enlisted to provide context on how unguided help-seeking behaviors evolve over a similar timeframe, rather than for direct statistical comparison.
In both settings, students used the JELAI environment during class assignments under their teacher's supervision. 
All participants and their parents provided informed consent.

The IG's course was divided into two six-week phases.
Participants were asked to complete two short surveys rating their previous experience with programming and with LLM chatbots.
In the pre-intervention phase, the unguided use of JELAI by students was logged, and a midterm exam was administered to identify baseline help-seeking behaviors and their correlation with performance.
Based on this initial analysis, we implemented a two-part pedagogical intervention. 
This included a class-wide discussion on the distinction between instrumental and executive help-seeking, followed by brief, one-on-one sessions where students reflected on their own query patterns with a researcher \cite{kozanitis_perception_2007}. 
In the post-intervention phase, students continued the course with JELAI access, and the intervention's effects were evaluated via logs and a final exam.

The CG used JELAI without any specific guidance for their entire course. 
The data collected included all interactions with the system and their final exam scores; no midterm exams or interventions were administered in this group.

All exams for both groups tested the concepts taught during the lessons, using a mix of syntax and coding tasks in JELAI, but with access to Juno deactivated.

\subsection{JELAI Environment and Data Classification}
The primary learning environment for both groups was JELAI \cite{valle_torre_jelai_2025}, an open-source platform that integrates an LLM-based chatbot, Juno, directly into a Jupyter Notebook interface (see Figure \ref{fig:JELAI}). 
Juno was powered by a Llama3.1:70b\footnote{\href{https://ollama.com/library/llama3.1}{https://ollama.com/library/llama3.1}} open-weights model, guided by a system prompt\footnote{\href{https://github.com/mvallet91/JELAI/blob/celdelta/history\_app.py\#L88}{https://github.com/mvallet91/JELAI/blob/celdelta/history\_app.py\#L88}} to act as a Socratic tutor that avoids providing direct solutions.
JELAI logged all student interactions—including chat messages, code cell executions, and errors—providing a fine-grained dataset of the learning process\footnote{The dataset is available at \href{https://data.4tu.nl/datasets/e947292a-bc0f-4da1-a160-58a7f4a49953/1}{https://data.4tu.nl/}}. 
These logs, along with student exam scores, served as the primary data for this study.

% JELAI utilizes Ollama to serve LLM models for interaction between students and the chatbot.
% For this study, Juno was based on Llama3.1:70b, the most capable open-weights model that could be loaded on our server (using an Nvida A40 GPU) at the time of this research, returning answers in 5-9 seconds\footnote{\href{https://ollama.com/library/llama3.1}{https://ollama.com/library/llama3.1}}. 
% Juno was prompted as a helpful AI tutor for introductory Python and instructed to avoid providing direct solutions, instead using the Socratic method to guide students to the answer.

To analyze student queries, we employed a multi-step classification process. 
Drawing from help-seeking theory \cite{nelson-le_gall_help-seeking_1981,ko_trees_2024}, we first manually coded the IG's pre-intervention queries into three high-level categories: Instrumental (seeking understanding), Executive (seeking solutions), and Other. 
This manually coded data was used to train and validate a few-shot LLM classifier using DSPy\footnote{\href{https://dspy.ai/}{https://dspy.ai/}}, which achieved an 84\% accuracy and was subsequently used to classify all queries from the CG. 
For a more detailed analysis within the IG, queries were also manually coded into finer-grained sub-categories \cite{sheese_patterns_2024,xiao_preliminary_2024} (see Figure \ref{fig:heatmap}).

\subsection{Data Analysis}
For RQ1, we used Spearman rank correlation ($\rho$) due to the small sample sizes of our groups and potential non-normality of the data. 
Similarly, for RQ2, we used Spearman rank correlation ($\rho$) to assess the relationship between instrumental/executive queries and the grades, as well as a detailed analysis per query type. 

For RQ3, investigating the intervention's influence, we employed non-parametric tests: the Wilcoxon signed-rank test was used to compare paired data (e.g., proportion of query types before and after the intervention for the same students, and midterm vs. final exam scores). 
All statistical tests were conducted with a significance level of $\alpha = 0.05$.

% Furthermore, we manually reviewed the interactions of several participants based on their performance and behavior changes.
Based on coding and dialog logs, we describe the learning activities of several participants to contextualize our findings and illuminate specific behaviors.
We also conduct sequence analysis on the entire group to further explore the behavioral changes.
We describe these insights in the Discussion (in Section \ref{sec:discussion}).

\section{Results}
This section presents the findings organized by our research questions, first establishing the baseline behaviors observed in both groups and then evaluating the impact of the intervention. 
Students in the Intervention Group (IG, N=18) submitted 480 queries to the LLM, with a mean of 25.4 queries per student. 
The Comparison Group (CG, N=19) submitted 403 queries, with a similar mean of 21.21 queries per user.

Based on pre-course survey for IG, the students were largely novice programmers, reporting a low average prior programming experience ($M=1.89, SD=1.02$) on a 5-point scale. 
Their self-reported prior usage of LLM chatbots was moderate, corresponding to an average use of 1-2 times per week ($M=3.06, SD=1.11$) on a 5-point scale.

% Regarding individual types, \texttt{Concept Comprehension} questions were the most frequent category, accounting for 24.7\% of all queries. 
% This highlights a significant need for students to understand programming concepts. 
% The second most common type was \texttt{Copying Notebook Questions} (19.2\%), which is concerning, as it suggests that a substantial portion of the students relied on the LLM to obtain direct answers rather than engage with the material, a behavior consistent with executive help-seeking.

% For the CG ...Number of users: 53
% Total messages: 1285
% User with least messages: 1
% User with most messages: 85
% Mean messages per user: 24.25
% Spearman correlation between total messages and total grade: -0.158 (p-value: 0.482)
% Spearman correlation between proportion of instrumental messages and total grade: -0.003 (p-value: 0.991)
% Spearman correlation between proportion of executive messages and total grade: -0.278 (p-value: 0.210)
% Spearman correlation between proportion of other messages and total grade: 0.302 (p-value: 0.172)
% For this group, we did not have questionnaires or subcategory classifications.
\begin{figure}
    \centering
    \includegraphics[width=0.95\linewidth]{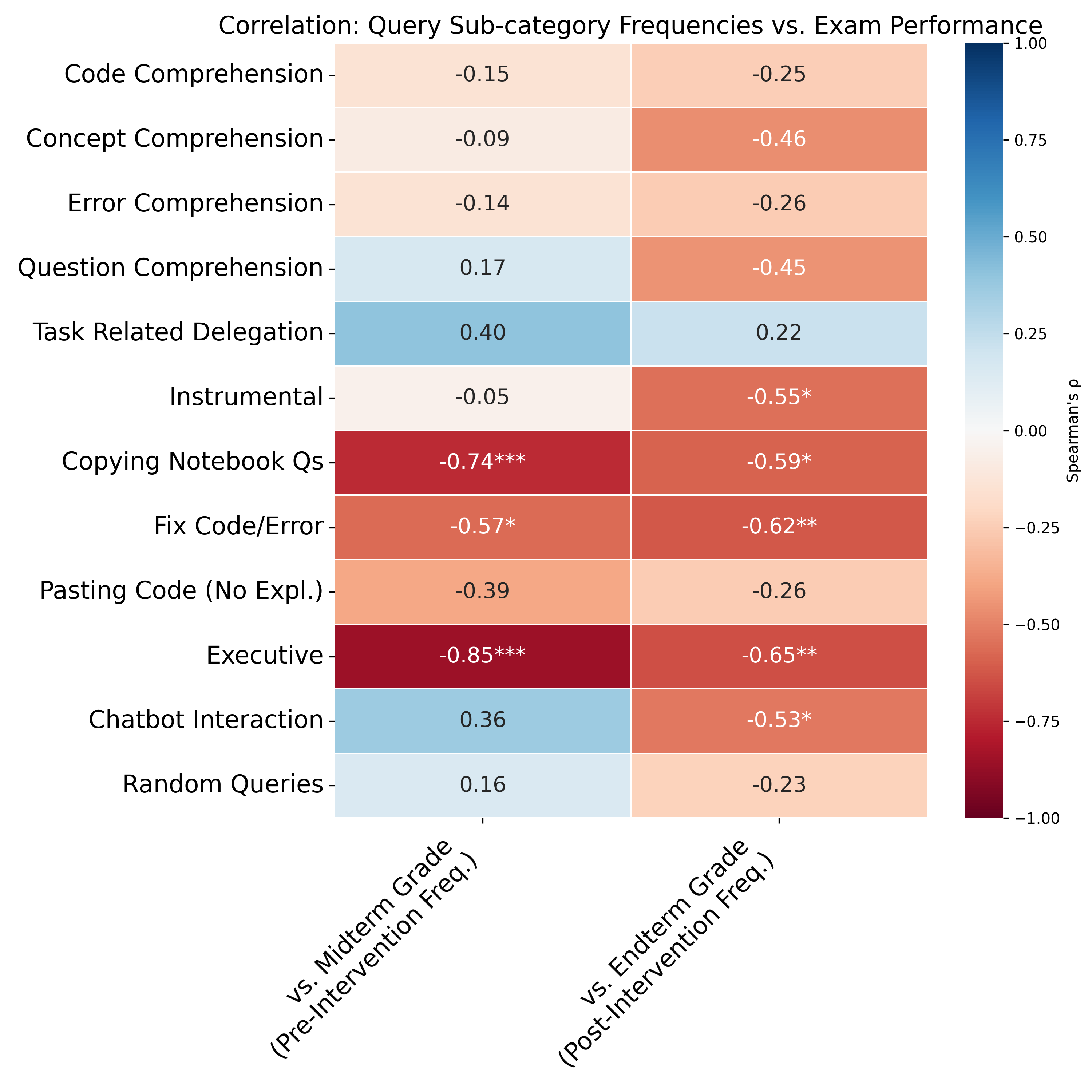}
    \caption{Correlations between student interaction types and grades, left column is data until the midterm correlated with the midterm grades, right column is the data after the midterm, correlated with the final exam grades. (* $p<0.05$, ** $p<0.01$, *** $p<0.001$)}
    \label{fig:heatmap}
\end{figure}

\textbf{RQ1: Correlation Between Interaction Frequency and Performance} 
In the IG's pre-intervention phase, we found a significant, negative correlation between the total number of LLM queries and midterm exam scores ($\rho = -0.502, p = 0.034$), indicating that higher interaction frequency was associated with lower performance. 
This is consistent with prior ITS research suggesting that frequency often signals a struggling student \cite{aleven_help_2016}. 
Further analysis suggested this was a selection effect; after controlling for students' self-reported prior programming experience, which showed a negative trend with interaction frequency ($\rho = -0.442, p = 0.067$), the direct relationship between frequency and grades became non-significant ($p = 0.084$). 
In the CG, no significant correlation was found between total query frequency and final exam grades ($\rho = -0.059, p = 0.811$)
\begin{figure*}
    \centering
    \includegraphics[width=0.75\linewidth]{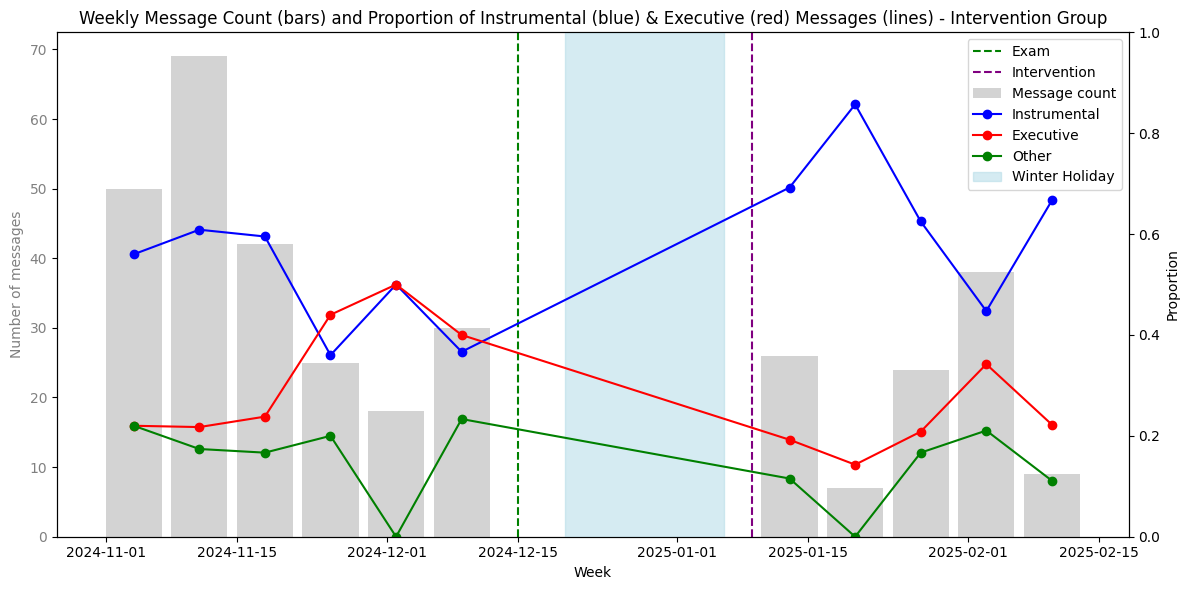}
    \caption{Weekly message counts and proportions per type for the Intervention Group}
    \label{fig:propcdl}
\end{figure*}

\begin{figure*}
    \centering
    \includegraphics[width=0.75\linewidth]{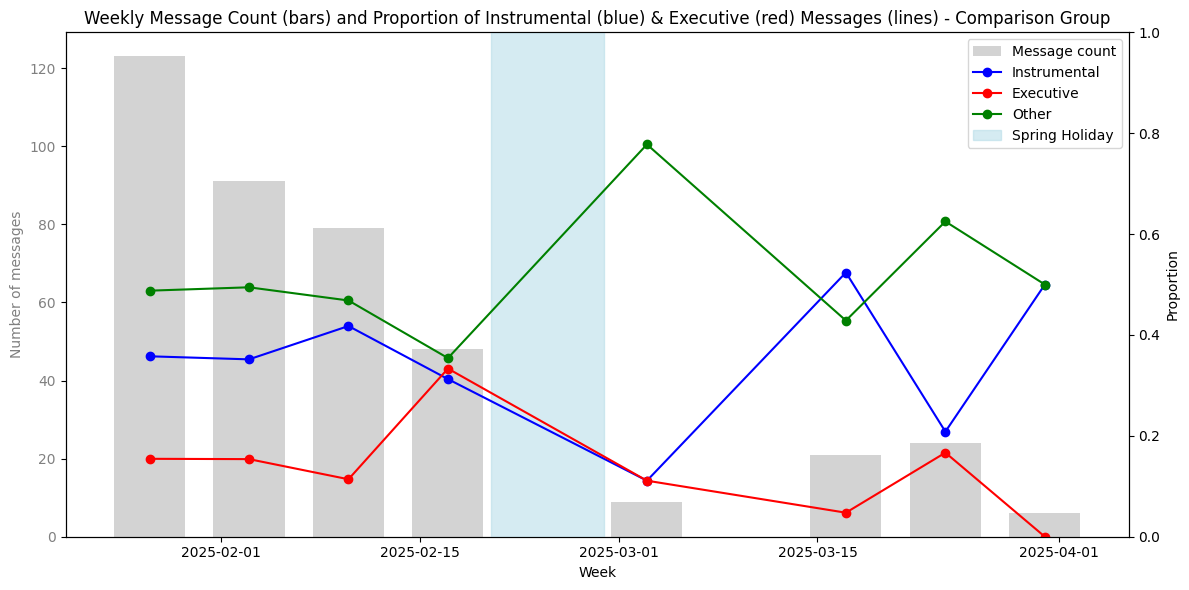}
    \caption{Weekly message counts and proportions per type for the Comparison Group}
    \label{fig:propstan}
\end{figure*}

\textbf{RQ2:  Impact of Instrumental vs. Executive Queries}
Analysis of query types in the IG's pre-intervention phase revealed the core problem for our DBR cycle. 
The most frequent query type was \texttt{Concept Comprehension} (24.7\%), followed by \texttt{Copying Notebook Questions} (19.2\%), a clear executive behavior. 
This was reflected in the correlations: there was a clear and significant negative correlation between the proportion of executive queries and midterm grades ($\rho = -0.851, p < 0.001$), while instrumental queries showed no significant correlation ($\rho = -0.052, p = 0.838$). 
The heatmap in Figure \ref{fig:heatmap} illustrates these relationships across all sub-categories.
In the CG, a similar directional trend was observed, with a weak, non-significant negative correlation between executive queries and final grades ($rho = -0.260, p = 0.281$). 
As shown in Figure \ref{fig:propstan}, this group's engagement declined after an initial novelty period. 
While the proportion of executive queries shows a downward trend over time, the pattern is marked by high fluctuation and sparse data in later weeks, preventing any firm conclusions about self-correction. 
This finding suggests that without guidance, unproductive help-seeking patterns do not reliably resolve on their own.

% \begin{figure*}
%     \centering
%     \includegraphics[width=0.9\textwidth]{figures/cdl-analysis-before-after-proportions.png}
%     \caption{Comparison of student query behavior before (left) and after (right) an intervention, positioned by midterm and endterm grades, respectively. 
%     Each bar is a student, with colors indicating the query type (Executive, Instrumental, Other). }
%     \label{fig:bars}
% \end{figure*}

% \textcolor{red}{CG tells us a different story: there was no correlation between grades and the number of queries or their category proportions.
% Students engaged in longer conversations with the chatbot (portrayed by the higher proportion of "Other" type messages in \ref{fig:propstan}).
% Furthermore, students gradually stop using Juno to help them and revert to asking their teachers for assistance, in line with the results from previous studies \cite{lyu_evaluating_2024}.}

\textbf{RQ3: Influence of the Pedagogical Intervention}
The intervention in the IG served as a clear inflection point, prompting a significant shift in behavior (Figure \ref{fig:propcdl}). 
A Wilcoxon signed-rank test on paired data (N = 14) showed that the proportion of executive queries, which had been rising, decreased significantly post-intervention, with a large effect size ($W = 15.0, p = 0.036, r = 0.560$). 
This behavioral change was accompanied by a reduction in overall query volume.
While the average final exam grade (M=7.56, SD=1.97) was higher than the midterm grade (M=7.28, SD=2.00), this improvement was not statistically significant ($W = 34.0, p = 0.083$), though the medium-to-large effect size ($r = 0.409$) suggests a potentially meaningful change.
This contrasts with the fluctuating, directionless patterns in the CG in Figure \ref{fig:propstan}, reinforcing that the observed shift in the IG was a direct result of the intervention.

\subsection{Qualitative Cases of LLM Interaction}
\label{sec:qualitative}
Qualitative analysis of the IG’s interaction logs provides crucial context for our findings, revealing distinct learner profiles and the nuanced impact of our intervention. 
High-achieving students demonstrated a variety of effective strategies. 
Some used the LLM sparingly, relying on strong independent skills, while others effectively used it as a conceptual partner, asking instrumental questions (e.g., “what is the difference between = and ==?”) instead of requesting direct fixes.

The intervention's potential is best illustrated by contrasting two students.
Mary, who initially relied on executive queries and pasting errors, shifted to asking for general algorithms after the intervention, leading to improved grades (7.30 → 8.65). 
Mary had a high number of total interactions with Juno and was one of the most active students in terms of code cell executions.
Conversely, Sam markedly improved their help-seeking strategy—shifting from 58\% executive queries to 64\% instrumental—but saw their grades decline (5.50 → 4.37). 
Sam's case critically suggests that while behavior can be changed, this may not be sufficient to overcome foundational knowledge gaps.

This behavioral shift was mirrored across the group. 
To understand changes in student workflows, we analyzed sequential patterns of log events using a fixed-window approach, examining the two actions preceding and two actions following each LLM interaction. 
Pre-intervention, a dominant pattern was reactive and superficial: students would encounter an error, immediately ask the AI for a fix, and then paste the provided code (\texttt{Edit $\rightarrow$ Error $\rightarrow$ Interaction $\rightarrow$ Edit $\rightarrow$ Success}). 
After the intervention, this pattern became less frequent. 
In its place, a more proactive and engaged workflow emerged, where students would successfully execute code and then consult the AI, likely for planning or conceptual clarification, before proceeding with another successful action (\texttt{Edit $\rightarrow$ Success $\rightarrow$ Interaction $\rightarrow$ Edit $\rightarrow$ Success}). 
This shift from using the AI as a reactive code-fixer to a proactive conceptual partner reinforces the positive behavioral change observed in the quantitative data.

\section{Discussion}
\label{sec:discussion}
Addressing RQ1, our finding of a significant negative correlation between unguided LLM usage and exam scores is consistent with prior research \cite{khurana_why_2024}. 
However, our analysis suggests this is not a simple causal relationship, as it contradicts results from similar studies with a more guided LLM programming tool \cite{sheese_patterns_2024,kazemitabaar_codeaid_2024}. 
A plausible explanation is a selection effect, whereby students with less prior programming experience—who are already more likely to struggle—are the ones self-selecting into higher usage of the tool \cite{aleven_help_2016}.
Controlling for the self-reported prior experience showed that the initial negative correlation is better explained by this selection effect.
This trend is echoed in our qualitative observations; the highest-performing students used the LLM sparingly, whereas students who struggled, such as Sam, demonstrated a higher query volume.

For RQ2, our analysis revealed a sharp contrast: executive help-seeking was detrimental to performance, whereas instrumental help-seeking had a more complex and non-significant correlation. 
This suggests that sustained high frequency of any help-seeking, even theoretically beneficial instrumental queries, likely indicates persistent difficulty that impacts overall performance \cite{marwan_unproductive_2020}. 
When a struggling student turns to the LLM, a tendency to seek direct solutions rather than explanations will naturally hinder learning and exam performance.
This pattern of superficial engagement is supported by our finding that 19.2\% of queries in the IG involved copying questions.
Such a pattern may then lead to a breakdown in the learning process, where the learner fails to actively process the information provided and just pastes the solution into the notebook \cite{wittwer_why_2008}.
Furthermore, the integration of the LLM may be changing the process from writing code to a more complex co-creation and verification, increasing the challenge for less experienced students \cite{tossell_student_2024}. 

The Comparison Group provides important context, illustrating that these unproductive patterns do not self-correct. 
Without guidance, the CG’s engagement with the chatbot was marked by fluctuating help-seeking proportions and an overall decline in use after an initial novelty period (Figure \ref{fig:propstan}). 
Instead of converging on productive strategies, students engaged in less-focused conversations, eventually diminishing the perceived usefulness of the tool and reverting to asking their human teacher for assistance—a finding consistent with prior work \cite{setala_use_2025,lyu_evaluating_2024}. 
This waning engagement likely explains the lack of a clear correlation between their usage and final grades.

Our mid-course intervention for RQ3 successfully guided students away from detrimental executive behaviors and toward more effective instrumental strategies \cite{ng_empowering_2024}. 
Post-intervention data indicated success in shifting relative usage: the percentage of executive questions decreased significantly for the group, with a large effect size. 
And while there was a reduction in use after the novelty of the first 3 weeks, it was not as marked as in the Comparison Group or similar studies \cite{lyu_evaluating_2024}.
This behavioral shift toward more instrumental approaches was also demonstrated in the post-intervention query patterns of both Mary and Sam (Section \ref{sec:qualitative}).

% Furthermore, our analysis of sequential pattern data suggested improved workflows: \texttt{Error $\rightarrow$ AI $\rightarrow$ Code Paste} cycles became less dominant, while patterns linking instrumental help to subsequent success (\texttt{Success $\rightarrow$ AI $\rightarrow$ Success}) appeared more frequently.
% While these behavioral shifts are positive, the observed overall grade increase from midterm to end-term was not statistically significant \cite{aleven_help_2016}.

\subsection{Practical Implications for Educators and System Designers}
The findings from this study provide several actionable recommendations for practitioners seeking to effectively integrate LLMs into programming education.

For Educators:
\begin{itemize}
    \item Teach Help-Seeking as a Skill: Do not assume students know how to use LLMs effectively. 
    Our intervention's success suggests that educators should explicitly teach the distinction between instrumental ("how does this concept work?") and executive ("fix my code") queries. 
    \item Design for Metacognition: Assignments should be designed to discourage simple solution-seeking. 
    Instead of tasks that can be solved with a single block of code, educators can create multi-part problems or require students to submit a brief "learning journal" with their code, explaining a key concept they learned from the LLM or a particularly helpful interaction. 
    This incentivizes and makes the learning process visible for students and instructors.
    \item Use Analytics for Formative Feedback: The methods used in this study—analyzing query types and error rates—can serve as a powerful tool for formative assessment. 
    An instructor could review interaction logs to identify groups of students who rely heavily on executive help and provide targeted, individual guidance long before a summative exam.
\end{itemize}

For System Designers:
\begin{itemize}
    \item Prioritize Pedagogy over Task Completion: Educational chatbots should not be neutral answer-bots. 
    The underlying system prompts must be engineered to be pedagogical agents that prioritize learning. 
    This means adding guardrails to the LLM to prevent it from providing direct solutions and instead guiding the student toward a conceptual understanding, much as a human tutor would.
    \item Implement Adaptive Scaffolding: Systems should be designed to detect patterns of unproductive learning behaviors, like executive help-seeking or help-avoidance. 
    For example, if a student submits consecutive "fix-it" queries, the system could automatically intervene with a metacognitive prompt, such as, "It looks like you're stuck, can you tell me what you were trying to do with this code?" 
    This automates the kind of guidance that proved effective in our intervention.
    \item Make Learning Visible to the Learner: Future iterations of educational programming environments could include a learner-facing dashboard that provides simple analytics (e.g., "This week, most of your questions were productive—great job focusing on improving your understanding!"). 
    This feedback loop empowers students to monitor and regulate their own learning behaviors, fostering the kind of self-awareness demonstrated by our high-achieving participants.
\end{itemize}

\subsection{Contributions}
This study makes several distinct contributions. 
Methodologically, it provides a concrete example of a DBR cycle applied to the integration of generative AI, demonstrating its value in moving from a vaguely defined problem ("students might misuse LLMs") to a specific, evidence-based diagnosis: the negative impact of unguided executive help-seeking. 
The iterative process of analysis, design, and evaluation refined our understanding, yielding tangible design principles rooted in authentic classroom practice.

Furthermore, the JELAI integration serves as a novel research instrument, capturing granular workflow data (e.g., the sequence of execution after a query) that is lost in studies relying on external chat interfaces.  
The analytical approach of classifying help-seeking intent (instrumental vs. executive) and analyzing error rates provides a replicable framework for future research.

Empirically, our exploratory study further strengthens evidence that simply providing access to an LLM does not guarantee positive learning outcomes and can, in fact, correlate with lower performance if students adopt non-adaptive strategies. 
Critically, the qualitative case studies—particularly the divergent outcomes of Mary and Sam—demonstrate that while help-seeking behaviors can be improved through direct intervention, overcoming foundational knowledge gaps presents a significant challenge.

Theoretically, this research extends help-seeking theory into the context of modern generative AI, confirming that the instrumental-executive framework is a powerful lens for understanding student behavior with these tools. 
By connecting the observed behaviors to established learning science concepts such as the "illusion of understanding" \cite{wittwer_why_2008}, this work highlights the critical need for educational LLMs to be designed not as answer providers, but as pedagogical tools that actively foster student metacognition and critical thinking \cite{ji_humanmachine_2025}.

\vspace{-0.15cm}
\subsection{Limitations}
This study has several limitations that warrant consideration. 
First, as an exploratory study, the small sample sizes restrict the generalizability of our findings. 
The observed negative correlation for specific instrumental comprehension subcategories (as shown in Figure \ref{fig:heatmap}) or the lack of significant correlation for total instrumental frequency may be heavily influenced by the specific usage patterns of the lowest- and highest-performing students.

Second, our analysis was confined to interactions with the integrated JELAI chatbot; unmonitored use of external LLMs (like ChatGPT) or other help sources (peers or family) could have confounded the relationship between measured usage and learning outcomes. 
While students were asked to only use Juno during class, there was no strict enforcement, and they were allowed to chat with each other or ask the teacher for help.

Third, the qualitative analysis of sequential patterns, while indicative, is based on aggregated top frequencies and does not capture workflow changes with rigor.
Finally, other individual factors, such as motivation and self-regulation, as well as variations in exam difficulty, may have also influenced the observed trends.

Finally, while this study focused on student help-seeking behaviors, it did not strictly evaluate the accuracy of the LLM’s responses. 
The risk of hallucinations or bias remains a critical challenge in educational AI, one that pedagogical interventions alone cannot resolve, necessitating further technical safeguards \cite{giannakos_promise_2024}.

\subsection{Future Work}
Future work should aim to address these limitations. 
Replicating this study with larger, more diverse samples is necessary to enhance generalizability and obtain more robust correlation estimates.
Research designs could incorporate methods to account for external tool usage and other forms of help-seeking, such as self-reporting or a more detailed log analysis. 

Further investigation into sequential patterns could involve more advanced sequence mining techniques on larger datasets to identify macro-trends, or potentially linking specific micro-patterns to learning outcomes at the individual student level \cite{valle_torre_sequence_2024}.
Furthermore, employing experimental designs that control for prior knowledge, motivation, and other individual differences would help isolate the specific impact of different LLM interaction types and frequencies.
Investigating the learning trajectories associated with specific help-seeking patterns, particularly for mid-range performers \cite{karumbaiah_context_2022}, and exploring optimal LLM design features that effectively scaffold learning remain important avenues for further research \cite{wu_ai_2024}.

\section{Conclusion}
This study investigated the role of LLMs in introductory high school programming, finding that they can act as both a learning partner and a hindrance. 
Our initial results confirm the risks: increased interaction, particularly driven by executive help-seeking, correlated with lower exam performance. 
While a targeted pedagogical intervention reduced detrimental behaviors, it did not lead to a statistically significant improvement in average grades.
Ultimately, this research concludes that the value of an LLM is not inherent to the technology but is critically dependent on pedagogy. 
Providing access alone is insufficient and can be counterproductive.
To ensure LLMs serve as a "friend" and not a "foe," educators and system designers must co-develop learning environments that explicitly teach and scaffold productive, instrumental learning strategies.

% \section*{Acknowledgments}
% This should be a simple paragraph before the References to thank those individuals and institutions who have supported your work on this article.

% {\appendix[JELAI System Prompt for Juno]

% }

% %% consistent spelling of the heading.
\begin{acks}
To Thom van der Velden and Stijn Risseeuw for their assistance in collecting data at the two schools.
% To the reviewers for their efforts.
\end{acks}

\printbibliography

@inproceedings{zambach_ai-enhanced_2025,
	address = {New York, NY, USA},
	series = {{SAC} '25},
	title = {{AI}-{Enhanced} {Learning}: {Comparing} {Outcomes} in {Introductory} and {Advanced} {Programming} {Courses}},
	isbn = {979-8-4007-0629-5},
	shorttitle = {{AI}-{Enhanced} {Learning}},
	url = {https://doi.org/10.1145/3672608.3707909},
	doi = {10.1145/3672608.3707909},
	urldate = {2025-10-09},
	booktitle = {Proceedings of the 40th {ACM}/{SIGAPP} {Symposium} on {Applied} {Computing}},
	publisher = {Association for Computing Machinery},
	author = {Zambach, Sine},
	month = may,
	year = {2025},
	pages = {104--105},
}

@inproceedings{setala_use_2025,
	address = {New York, NY, USA},
	series = {{SAC} '25},
	title = {The {Use} of {Generative} {Artificial} {Intelligence} for {Upper} {Secondary} {Mathematics} {Education} {Through} the {Lens} of {Technology} {Acceptance}},
	isbn = {979-8-4007-0629-5},
	url = {https://dl.acm.org/doi/10.1145/3672608.3707817},
	doi = {10.1145/3672608.3707817},
	urldate = {2025-10-09},
	booktitle = {Proceedings of the 40th {ACM}/{SIGAPP} {Symposium} on {Applied} {Computing}},
	publisher = {Association for Computing Machinery},
	author = {Setälä, Mika and Heilala, Ville and Sikström, Pieta and Kärkkäinen, Tommi},
	month = may,
	year = {2025},
	pages = {74--82},
}

@article{wittwer_why_2008,
	title = {Why {Instructional} {Explanations} {Often} {Do} {Not} {Work}: {A} {Framework} for {Understanding} the {Effectiveness} of {Instructional} {Explanations}},
	volume = {43},
	issn = {0046-1520},
	shorttitle = {Why {Instructional} {Explanations} {Often} {Do} {Not} {Work}},
	url = {https://doi.org/10.1080/00461520701756420},
	doi = {10.1080/00461520701756420},
	number = {1},
	urldate = {2025-06-02},
	journal = {Educational Psychologist},
	author = {Wittwer, Jörg and Renkl, Alexander},
	month = jan,
	year = {2008},
	note = {Publisher: Routledge
\_eprint: https://doi.org/10.1080/00461520701756420},
	pages = {49--64},
}

@article{kozanitis_perception_2007,
	title = {Perception of {Teacher} {Support} and {Reaction} towards {Questioning}: {Its} {Relation} to {Instrumental} {Help}-{Seeking} and {Motivation} to {Learn}},
	volume = {19},
	issn = {1812-9129},
	shorttitle = {Perception of {Teacher} {Support} and {Reaction} towards {Questioning}},
	url = {https://eric.ed.gov/?id=EJ901297},
	language = {en},
	number = {3},
	urldate = {2025-10-07},
	journal = {International Journal of Teaching and Learning in Higher Education},
	author = {Kozanitis, Anastassis and Desbiens, Jean-Francois and Chouinard, Roch},
	year = {2007},
	note = {Publisher: International Society for Exploring Teaching and Learning
ERIC Number: EJ901297},
	pages = {238--250},
}

@inproceedings{valle_torre_jelai_2025,
	address = {Cham},
	title = {{JELAI}: {Integrating} {AI} and {Learning} {Analytics} in {Jupyter} {Notebooks}},
	isbn = {978-3-031-98465-5},
	shorttitle = {{JELAI}},
	doi = {10.1007/978-3-031-98465-5_9},
	language = {en},
	booktitle = {Artificial {Intelligence} in {Education}},
	publisher = {Springer Nature Switzerland},
	author = {Valle Torre, Manuel and van der Velden, Thom and Specht, Marcus and Oertel, Catharine},
	editor = {Cristea, Alexandra I. and Walker, Erin and Lu, Yu and Santos, Olga C. and Isotani, Seiji},
	year = {2025},
	pages = {68--75},
}

@inproceedings{park_empowering_2024,
	address = {New York, NY, USA},
	series = {{CHI} {EA} '24},
	title = {Empowering {Personalized} {Learning} through a {Conversation}-based {Tutoring} {System} with {Student} {Modeling}},
	isbn = {979-8-4007-0331-7},
	url = {https://dl.acm.org/doi/10.1145/3613905.3651122},
	doi = {10.1145/3613905.3651122},
	urldate = {2025-09-18},
	booktitle = {Extended {Abstracts} of the {CHI} {Conference} on {Human} {Factors} in {Computing} {Systems}},
	publisher = {Association for Computing Machinery},
	author = {Park, Minju and Kim, Sojung and Lee, Seunghyun and Kwon, Soonwoo and Kim, Kyuseok},
	month = may,
	year = {2024},
	pages = {1--10},
}

@article{tossell_student_2024,
	title = {Student {Perceptions} of {ChatGPT} {Use} in a {College} {Essay} {Assignment}: {Implications} for {Learning}, {Grading}, and {Trust} in {Artificial} {Intelligence}},
	volume = {17},
	issn = {1939-1382},
	shorttitle = {Student {Perceptions} of {ChatGPT} {Use} in a {College} {Essay} {Assignment}},
	url = {https://ieeexplore.ieee.org/document/10400910/},
	doi = {10.1109/TLT.2024.3355015},
	urldate = {2025-06-19},
	journal = {IEEE Transactions on Learning Technologies},
	author = {Tossell, Chad C. and Tenhundfeld, Nathan L. and Momen, Ali and Cooley, Katrina and de Visser, Ewart J.},
	year = {2024},
	pages = {1069--1081},
}

@article{ji_humanmachine_2025,
	title = {Human–{Machine} {Cocreation}: {The} {Effects} of {ChatGPT} on {Students}’ {Learning} {Performance}, {AI} {Awareness}, {Critical} {Thinking}, and {Cognitive} {Load} in a {STEM} {Course} {Toward} {Entrepreneurship}},
	volume = {18},
	issn = {1939-1382},
	shorttitle = {Human–{Machine} {Cocreation}},
	url = {https://ieeexplore.ieee.org/document/10938864/},
	doi = {10.1109/TLT.2025.3554584},
	urldate = {2025-06-19},
	journal = {IEEE Transactions on Learning Technologies},
	author = {Ji, Yu and Zhan, Zehui and Li, Tingting and Zou, Xuanxuan and Lyu, Siyuan},
	year = {2025},
	pages = {402--415},
}

@article{yang_systematic_2023,
	title = {A {Systematic} {Review} of {Studies} {Exploring} {Help}-{Seeking} {Strategies} in {Online} {Learning} {Environments}},
	volume = {27},
	copyright = {https://creativecommons.org/licenses/by/4.0/},
	issn = {2472-5730, 2472-5749},
	url = {https://olj.onlinelearningconsortium.org/index.php/olj/article/view/3400},
	doi = {10.24059/olj.v27i1.3400},
	language = {en},
	number = {1},
	urldate = {2024-11-20},
	journal = {Online Learning},
	author = {Yang, Fan and Stefaniak, Jill},
	month = mar,
	year = {2023},
}

@article{giannakos_promise_2024,
	title = {The promise and challenges of generative {AI} in education},
	volume = {0},
	issn = {0144-929X},
	url = {https://doi.org/10.1080/0144929X.2024.2394886},
	doi = {10.1080/0144929X.2024.2394886},
	number = {0},
	urldate = {2025-01-30},
	journal = {Behaviour \& Information Technology},
	author = {Giannakos, Michail and Azevedo, Roger and Brusilovsky, Peter and Cukurova, Mutlu and Dimitriadis, Yannis and Hernandez-Leo, Davinia and Järvelä, Sanna and Mavrikis, Manolis and Rienties, Bart},
	year = {2024},
	pages = {1--27},
}

@article{fan_beware_2024,
	title = {Beware of metacognitive laziness: {Effects} of generative artificial intelligence on learning motivation, processes, and performance},
	volume = {n/a},
	copyright = {© 2024 British Educational Research Association.},
	issn = {1467-8535},
	shorttitle = {Beware of metacognitive laziness},
	url = {https://onlinelibrary.wiley.com/doi/abs/10.1111/bjet.13544},
	doi = {10.1111/bjet.13544},
	language = {en},
	number = {n/a},
	urldate = {2025-01-20},
	journal = {British Journal of Educational Technology},
	author = {Fan, Yizhou and Tang, Luzhen and Le, Huixiao and Shen, Kejie and Tan, Shufang and Zhao, Yueying and Shen, Yuan and Li, Xinyu and Gašević, Dragan},
	month = dec,
	year = {2024},
}

@inproceedings{lyu_evaluating_2024,
	address = {New York, NY, USA},
	series = {L@{S} '24},
	title = {Evaluating the {Effectiveness} of {LLMs} in {Introductory} {Computer} {Science} {Education}: {A} {Semester}-{Long} {Field} {Study}},
	isbn = {979-8-4007-0633-2},
	shorttitle = {Evaluating the {Effectiveness} of {LLMs} in {Introductory} {Computer} {Science} {Education}},
	url = {https://dl.acm.org/doi/10.1145/3657604.3662036},
	doi = {10.1145/3657604.3662036},
	urldate = {2025-03-25},
	booktitle = {Proceedings of the {Eleventh} {ACM} {Conference} on {Learning} @ {Scale}},
	publisher = {Association for Computing Machinery},
	author = {Lyu, Wenhan and Wang, Yimeng and Chung, Tingting (Rachel) and Sun, Yifan and Zhang, Yixuan},
	month = jul,
	year = {2024},
	pages = {63--74},
}

@inproceedings{price_factors_2017,
	address = {New York, NY, USA},
	series = {{ICER} '17},
	title = {Factors {Influencing} {Students}' {Help}-{Seeking} {Behavior} while {Programming} with {Human} and {Computer} {Tutors}},
	isbn = {978-1-4503-4968-0},
	url = {https://dl.acm.org/doi/10.1145/3105726.3106179},
	doi = {10.1145/3105726.3106179},
	urldate = {2025-03-18},
	booktitle = {Proceedings of the 2017 {ACM} {Conference} on {International} {Computing} {Education} {Research}},
	publisher = {Association for Computing Machinery},
	author = {Price, Thomas W. and Liu, Zhongxiu and Cateté, Veronica and Barnes, Tiffany},
	month = aug,
	year = {2017},
	pages = {127--135},
}

@article{zhai_effects_2024,
	title = {The effects of over-reliance on {AI} dialogue systems on students' cognitive abilities: a systematic review},
	volume = {11},
	issn = {2196-7091},
	shorttitle = {The effects of over-reliance on {AI} dialogue systems on students' cognitive abilities},
	url = {https://doi.org/10.1186/s40561-024-00316-7},
	doi = {10.1186/s40561-024-00316-7},
	number = {1},
	urldate = {2025-03-14},
	journal = {Smart Learning Environments},
	author = {Zhai, Chunpeng and Wibowo, Santoso and Li, Lily D.},
	month = jun,
	year = {2024},
	pages = {28},
}

@article{cheah_factors_2020,
	title = {Factors {Contributing} to the {Difficulties} in {Teaching} and {Learning} of {Computer} {Programming}: {A} {Literature} {Review}},
	volume = {12},
	shorttitle = {Factors {Contributing} to the {Difficulties} in {Teaching} and {Learning} of {Computer} {Programming}},
	url = {https://eric.ed.gov/?id=EJ1267658},
	language = {en},
	number = {2},
	urldate = {2025-03-14},
	journal = {Contemporary Educational Technology},
	author = {Cheah, Chin Soon},
	year = {2020},
	note = {ERIC Number: EJ1267658},
}

@inproceedings{pal_chowdhury_autotutor_2024,
	address = {New York, NY, USA},
	series = {L@{S} '24},
	title = {{AutoTutor} meets {Large} {Language} {Models}: {A} {Language} {Model} {Tutor} with {Rich} {Pedagogy} and {Guardrails}},
	isbn = {9798400706332},
	shorttitle = {{AutoTutor} meets {Large} {Language} {Models}},
	url = {https://dl.acm.org/doi/10.1145/3657604.3662041},
	doi = {10.1145/3657604.3662041},
	urldate = {2024-09-26},
	booktitle = {Proceedings of the {Eleventh} {ACM} {Conference} on {Learning} @ {Scale}},
	publisher = {Association for Computing Machinery},
	author = {Pal Chowdhury, Sankalan and Zouhar, Vilém and Sachan, Mrinmaya},
	month = jul,
	year = {2024},
	pages = {5--15},
}

@article{karumbaiah_context_2022,
	title = {Context {Matters}: {Differing} {Implications} of {Motivation} and {Help}-{Seeking} in {Educational} {Technology}},
	volume = {32},
	issn = {1560-4306},
	shorttitle = {Context {Matters}},
	url = {https://doi.org/10.1007/s40593-021-00272-0},
	doi = {10.1007/s40593-021-00272-0},
	language = {en},
	number = {3},
	urldate = {2025-01-31},
	journal = {International Journal of Artificial Intelligence in Education},
	author = {Karumbaiah, Shamya and Ocumpaugh, Jaclyn and Baker, Ryan S.},
	month = sep,
	year = {2022},
	pages = {685--724},
}

@article{karabenick_understanding_2011,
	title = {Understanding and facilitating self-regulated help seeking},
	volume = {2011},
	copyright = {Copyright © 2011 Wiley Periodicals, Inc., A Wiley Company},
	issn = {1536-0768},
	url = {https://onlinelibrary.wiley.com/doi/abs/10.1002/tl.442},
	doi = {10.1002/tl.442},
	language = {en},
	number = {126},
	urldate = {2025-01-29},
	journal = {New Directions for Teaching and Learning},
	author = {Karabenick, Stuart A. and Dembo, Myron H.},
	year = {2011},
	note = {\_eprint: https://onlinelibrary.wiley.com/doi/pdf/10.1002/tl.442},
	pages = {33--43},
}

@inproceedings{becker_programming_2023,
	address = {Toronto ON Canada},
	title = {Programming {Is} {Hard} - {Or} at {Least} {It} {Used} to {Be}: {Educational} {Opportunities} and {Challenges} of {AI} {Code} {Generation}},
	isbn = {978-1-4503-9431-4},
	shorttitle = {Programming {Is} {Hard} - {Or} at {Least} {It} {Used} to {Be}},
	url = {https://dl.acm.org/doi/10.1145/3545945.3569759},
	doi = {10.1145/3545945.3569759},
	language = {en},
	urldate = {2025-01-30},
	booktitle = {Proceedings of the 54th {ACM} {Technical} {Symposium} on {Computer} {Science} {Education} {V}. 1},
	publisher = {ACM},
	author = {Becker, Brett A. and Denny, Paul and Finnie-Ansley, James and Luxton-Reilly, Andrew and Prather, James and Santos, Eddie Antonio},
	month = mar,
	year = {2023},
	pages = {500--506},
}

@article{nelson-le_gall_help-seeking_1981,
	title = {Help-seeking: {An} understudied problem-solving skill in children},
	volume = {1},
	issn = {0273-2297},
	shorttitle = {Help-seeking},
	url = {https://www.sciencedirect.com/science/article/pii/0273229781900198},
	doi = {10.1016/0273-2297(81)90019-8},
	number = {3},
	urldate = {2025-01-21},
	journal = {Developmental Review},
	author = {Nelson-Le Gall, Sharon},
	month = sep,
	year = {1981},
	pages = {224--246},
}

@article{ng_empowering_2024,
	title = {Empowering student self-regulated learning and science education through {ChatGPT}: {A} pioneering pilot study},
	volume = {55},
	copyright = {© 2024 The Authors. British Journal of Educational Technology published by John Wiley \& Sons Ltd on behalf of British Educational Research Association.},
	issn = {1467-8535},
	shorttitle = {Empowering student self-regulated learning and science education through {ChatGPT}},
	url = {https://onlinelibrary.wiley.com/doi/abs/10.1111/bjet.13454},
	doi = {10.1111/bjet.13454},
	language = {en},
	number = {4},
	urldate = {2025-01-20},
	journal = {British Journal of Educational Technology},
	author = {Ng, Davy Tsz Kit and Tan, Chee Wei and Leung, Jac Ka Lok},
	year = {2024},
	note = {\_eprint: https://onlinelibrary.wiley.com/doi/pdf/10.1111/bjet.13454},
	pages = {1328--1353},
}

@article{wu_ai_2024,
	title = {Do {AI} chatbots improve students learning outcomes? {Evidence} from a meta-analysis},
	volume = {55},
	copyright = {© 2023 British Educational Research Association.},
	issn = {1467-8535},
	shorttitle = {Do {AI} chatbots improve students learning outcomes?},
	url = {https://onlinelibrary.wiley.com/doi/abs/10.1111/bjet.13334},
	doi = {10.1111/bjet.13334},
	language = {en},
	number = {1},
	urldate = {2025-01-20},
	journal = {British Journal of Educational Technology},
	author = {Wu, Rong and Yu, Zhonggen},
	year = {2024},
	note = {\_eprint: https://onlinelibrary.wiley.com/doi/pdf/10.1111/bjet.13334},
	pages = {10--33},
}

@article{chuang_analyzing_2024,
	title = {Analyzing novice and competent programmers' problem-solving behaviors using an automated evaluation system},
	volume = {237},
	issn = {0167-6423},
	url = {https://www.sciencedirect.com/science/article/pii/S0167642324000613},
	doi = {10.1016/j.scico.2024.103138},
	urldate = {2025-01-16},
	journal = {Science of Computer Programming},
	author = {Chuang, Yung-Ting and Chang, Hsin-Yu},
	month = oct,
	year = {2024},
	pages = {103138},
}

@article{roll_benefits_2014,
	title = {On the {Benefits} of {Seeking} (and {Avoiding}) {Help} in {Online} {Problem}-{Solving} {Environments}},
	volume = {23},
	issn = {1050-8406},
	url = {https://doi.org/10.1080/10508406.2014.883977},
	doi = {10.1080/10508406.2014.883977},
	number = {4},
	urldate = {2024-11-28},
	journal = {Journal of the Learning Sciences},
	author = {Roll, Ido and Baker, Ryan S. J. d. and Aleven, Vincent and Koedinger, Kenneth R.},
	month = oct,
	year = {2014},
	note = {Publisher: Routledge
\_eprint: https://doi.org/10.1080/10508406.2014.883977},
	pages = {537--560},
}

@article{zhang_systematic_2024,
	title = {A {Systematic} {Literature} {Review} of {Empirical} {Research} on {Applying} {Generative} {Artificial} {Intelligence} in {Education}},
	volume = {1},
	issn = {2097-3926},
	url = {https://doi.org/10.1007/s44366-024-0028-5},
	doi = {10.1007/s44366-024-0028-5},
	language = {en},
	number = {3},
	urldate = {2024-11-11},
	journal = {Frontiers of Digital Education},
	author = {Zhang, Xin and Zhang, Peng and Shen, Yuan and Liu, Min and Wang, Qiong and Gašević, Dragan and Fan, Yizhou},
	month = sep,
	year = {2024},
	pages = {223--245},
}

@inproceedings{xie_developing_2023,
	address = {New York, NY, USA},
	series = {{ICER} '23},
	title = {Developing {Novice} {Programmers}’ {Self}-{Regulation} {Skills} with {Code} {Replays}},
	volume = {1},
	isbn = {978-1-4503-9976-0},
	url = {https://dl.acm.org/doi/10.1145/3568813.3600127},
	doi = {10.1145/3568813.3600127},
	urldate = {2024-11-08},
	booktitle = {Proceedings of the 2023 {ACM} {Conference} on {International} {Computing} {Education} {Research} - {Volume} 1},
	publisher = {Association for Computing Machinery},
	author = {Xie, Benjamin and Lim, Jared Ordona and Pham, Paul K.D. and Li, Min and Ko, Amy J.},
	month = sep,
	year = {2023},
	pages = {298--313},
}

@article{aleven_example-tracing_2016,
	title = {Example-{Tracing} {Tutors}: {Intelligent} {Tutor} {Development} for {Non}-programmers},
	volume = {26},
	issn = {1560-4306},
	shorttitle = {Example-{Tracing} {Tutors}},
	url = {https://doi.org/10.1007/s40593-015-0088-2},
	doi = {10.1007/s40593-015-0088-2},
	language = {en},
	number = {1},
	urldate = {2024-11-01},
	journal = {International Journal of Artificial Intelligence in Education},
	author = {Aleven, Vincent and McLaren, Bruce M. and Sewall, Jonathan and van Velsen, Martin and Popescu, Octav and Demi, Sandra and Ringenberg, Michael and Koedinger, Kenneth R.},
	month = mar,
	year = {2016},
	pages = {224--269},
}

@inproceedings{kazemitabaar_codeaid_2024,
	address = {New York, NY, USA},
	series = {{CHI} '24},
	title = {{CodeAid}: {Evaluating} a {Classroom} {Deployment} of an {LLM}-based {Programming} {Assistant} that {Balances} {Student} and {Educator} {Needs}},
	isbn = {9798400703300},
	shorttitle = {{CodeAid}},
	url = {https://dl.acm.org/doi/10.1145/3613904.3642773},
	doi = {10.1145/3613904.3642773},
	urldate = {2024-11-01},
	booktitle = {Proceedings of the 2024 {CHI} {Conference} on {Human} {Factors} in {Computing} {Systems}},
	publisher = {Association for Computing Machinery},
	author = {Kazemitabaar, Majeed and Ye, Runlong and Wang, Xiaoning and Henley, Austin Zachary and Denny, Paul and Craig, Michelle and Grossman, Tovi},
	month = may,
	year = {2024},
	pages = {1--20},
}

@inproceedings{sheese_patterns_2024,
	address = {New York, NY, USA},
	series = {{ACE} '24},
	title = {Patterns of {Student} {Help}-{Seeking} {When} {Using} a {Large} {Language} {Model}-{Powered} {Programming} {Assistant}},
	isbn = {9798400716195},
	url = {https://dl.acm.org/doi/10.1145/3636243.3636249},
	doi = {10.1145/3636243.3636249},
	urldate = {2024-10-24},
	booktitle = {Proceedings of the 26th {Australasian} {Computing} {Education} {Conference}},
	publisher = {Association for Computing Machinery},
	author = {Sheese, Brad and Liffiton, Mark and Savelka, Jaromir and Denny, Paul},
	month = jan,
	year = {2024},
	pages = {49--57},
}

@inproceedings{ko_trees_2024,
	address = {New York, NY, USA},
	series = {{ICER} '24},
	title = {The {Trees} in the {Forest}: {Characterizing} {Computing} {Students}' {Individual} {Help}-{Seeking} {Approaches}},
	volume = {1},
	isbn = {9798400704758},
	shorttitle = {The {Trees} in the {Forest}},
	url = {https://dl.acm.org/doi/10.1145/3632620.3671099},
	doi = {10.1145/3632620.3671099},
	urldate = {2024-10-28},
	booktitle = {Proceedings of the 2024 {ACM} {Conference} on {International} {Computing} {Education} {Research} - {Volume} 1},
	publisher = {Association for Computing Machinery},
	author = {Ko, Shao-Heng and Stephens-Martinez, Kristin},
	month = aug,
	year = {2024},
	pages = {343--358},
}

@inproceedings{xiao_preliminary_2024,
	address = {Atlanta, GA},
	title = {A {Preliminary} {Analysis} of {Students}' {Help} {Requests} with an {LLM}-powered {Chatbot} when {Completing} {CS1} {Assignments}},
	language = {en},
	booktitle = {8th {Educational} {Data} {Mining} in {Computer} {Science} {Education} {Workshop}},
	author = {Xiao, Ruiwei and Hou, Xinying and Kumar, Harsh and Moore, Steven and Stamper, John and Liut, Michael},
	month = jul,
	year = {2024},
}

@article{aleven_help_2016,
	title = {Help {Helps}, {But} {Only} {So} {Much}: {Research} on {Help} {Seeking} with {Intelligent} {Tutoring} {Systems}},
	volume = {26},
	issn = {1560-4306},
	shorttitle = {Help {Helps}, {But} {Only} {So} {Much}},
	url = {https://doi.org/10.1007/s40593-015-0089-1},
	doi = {10.1007/s40593-015-0089-1},
	language = {en},
	number = {1},
	urldate = {2024-05-28},
	journal = {International Journal of Artificial Intelligence in Education},
	author = {Aleven, Vincent and Roll, Ido and McLaren, Bruce M. and Koedinger, Kenneth R.},
	month = mar,
	year = {2016},
	pages = {205--223},
}

@inproceedings{marwan_unproductive_2020,
	address = {New York, NY, USA},
	series = {{ITiCSE} '20},
	title = {Unproductive {Help}-seeking in {Programming}: {What} it is and {How} to {Address} it},
	isbn = {978-1-4503-6874-2},
	shorttitle = {Unproductive {Help}-seeking in {Programming}},
	url = {https://dl.acm.org/doi/10.1145/3341525.3387394},
	doi = {10.1145/3341525.3387394},
	urldate = {2024-05-28},
	booktitle = {Proceedings of the 2020 {ACM} {Conference} on {Innovation} and {Technology} in {Computer} {Science} {Education}},
	publisher = {Association for Computing Machinery},
	author = {Marwan, Samiha and Dombe, Anay and Price, Thomas W.},
	month = jun,
	year = {2020},
	pages = {54--60},
}

@inproceedings{hou_effects_2024,
	address = {New York, NY, USA},
	series = {{ACE} '24},
	title = {The {Effects} of {Generative} {AI} on {Computing} {Students}’ {Help}-{Seeking} {Preferences}},
	isbn = {9798400716195},
	url = {https://dl.acm.org/doi/10.1145/3636243.3636248},
	doi = {10.1145/3636243.3636248},
	urldate = {2024-05-28},
	booktitle = {Proceedings of the 26th {Australasian} {Computing} {Education} {Conference}},
	publisher = {Association for Computing Machinery},
	author = {Hou, Irene and Mettille, Sophia and Man, Owen and Li, Zhuo and Zastudil, Cynthia and MacNeil, Stephen},
	month = jan,
	year = {2024},
	pages = {39--48},
}

@inproceedings{khurana_why_2024,
	address = {New York, NY, USA},
	series = {{IUI} '24},
	title = {Why and {When} {LLM}-{Based} {Assistants} {Can} {Go} {Wrong}: {Investigating} the {Effectiveness} of {Prompt}-{Based} {Interactions} for {Software} {Help}-{Seeking}},
	isbn = {9798400705083},
	shorttitle = {Why and {When} {LLM}-{Based} {Assistants} {Can} {Go} {Wrong}},
	url = {https://dl.acm.org/doi/10.1145/3640543.3645200},
	doi = {10.1145/3640543.3645200},
	urldate = {2024-05-28},
	booktitle = {Proceedings of the 29th {International} {Conference} on {Intelligent} {User} {Interfaces}},
	publisher = {Association for Computing Machinery},
	author = {Khurana, Anjali and Subramonyam, Hariharan and Chilana, Parmit K},
	month = apr,
	year = {2024},
	pages = {288--303},
}

@inproceedings{valle_torre_sequence_2024,
	address = {New York, NY, USA},
	series = {{LAK} '24},
	title = {The {Sequence} {Matters} in {Learning} - {A} {Systematic} {Literature} {Review}},
	isbn = {9798400716188},
	url = {https://doi.org/10.1145/3636555.3636880},
	doi = {10.1145/3636555.3636880},
	urldate = {2024-03-10},
	booktitle = {Proceedings of the 14th {Learning} {Analytics} and {Knowledge} {Conference}},
	publisher = {Association for Computing Machinery},
	author = {Valle Torre, Manuel and Oertel, Catharine and Specht, Marcus},
	month = mar,
	year = {2024},
	pages = {263--272},
}

@String{Computing = "Computing" }

@String{Computer = "{IEEE} Computer" }

@String{Springer = "Springer-Verlag" }

\end{document}